\def\year{2017}\relax
\documentclass[letterpaper]{article}
\usepackage{aaai16}
\usepackage{times}
\usepackage{helvet}
\usepackage{courier}
\usepackage{float}
\usepackage{graphicx}
\usepackage{enumerate}
\usepackage[usenames, dvipsnames]{color}
\usepackage{comment}

\begin{document}
%
\title{On Quitting: Performance and Practice in Online Game Play}
\author{
Tushar Agarwal\\
Indian Institute of Technology Ropar\\
Punjab, India 140001\\
\texttt{tushar.agarwal@iitrpr.ac.in}\\
\And
Keith Burghardt\\
University of California at Davis\\
Davis, CA 95616\\
\texttt{kaburghardt@ucdavis.edu}\\
\And
Kristina Lerman\\
USC Information Sciences Institute\\
Marina del Rey, CA 90292 \\
\texttt{lerman@isi.edu}
}

\maketitle
\begin{abstract}
We study the relationship between performance and practice by analyzing the activity of many players of a casual online game. We find significant heterogeneity in the improvement of player performance, given by score, and address this by dividing players into similar skill levels and segmenting each player's activity into sessions, i.e., sequence of game rounds without an extended break. After disaggregating data, we find that performance improves with practice across all skill levels. More interestingly, players are more likely to end their session after an especially large improvement, leading to a peak score in their very last game of a session. In addition, success is strongly correlated with a lower quitting rate when the score drops, and only weakly correlated with skill, in line with psychological findings about the value of persistence and ``grit'': successful players are those who persist in their practice despite lower scores.
Finally, we train an $\epsilon$-machine, a type of hidden Markov model, and find a plausible mechanism of game play that can predict player performance and quitting the game.
Our work raises the possibility of real-time assessment and behavior prediction that can be used to optimize human performance.

\end{abstract}

\section{Introduction}
\begin{quote}
\emph{
How much grit do you think you've got?\\
Can you quit a thing that you like a lot?\\
} -- ``On Quitting'' by Edgar Guest
\end{quote}

How do people achieve mastery? What distinguishes high achievers from average performers?
Performance generally improves with practice, as demonstrated on a variety of tasks in the laboratory setting and in the field~\cite{newell1981mechanisms}, suggesting that with enough practice even mediocre performers can approach the mastery of successful individuals.
However, not all practice is equally effective in helping achieve mastery. 
Deliberate practice, which emphasizes quality, not quantity of practice, improves performance most~\cite{Duckworth2011deliberate,ericsson2006influence}. The search for individual traits responsible for variations in the capacity for deliberate practice uncovered \emph{grit}, a trait related to psychological constructs, such as persistence, resilience, and self-control, which enables individuals to persevere in their efforts to achieve their goals
~\cite{duckworth2007grit}. 
Grit may explain the self-discipline to continue practicing, 
even when faced with temporary setbacks, such as a short-term drop in performance.


Recent proliferation of behavioral data collected ``in the wild'' enables longitudinal studies to explore and validate these findings. 
We carry out an empirical analysis of online game play to quantify individual traits associated with success.
The data we study consists of records of over 850K players of a game called Axon. Following \cite{stafford2014tracing}, who first studied this data, we operationalize performance as player's score, and practice as playing rounds of the game.
Like other behavioral data, Axon data presents analytic challenges. It is extremely \emph{noisy}: requiring aggregating variables over the population. It is also \emph{heterogeneous}: composed of differently-behaving subgroups varying in size according to the Pareto distribution. As a result, the trends observed in aggregated data may be quite different from those of the underlying subgroups~\cite{Vaupel85heterogeneity}.
To address this effect, known as Simpson's paradox, we disaggregate data by user skill 
and activity. 
We segment player activity into sessions, where a session is a sequence of games without an extended break. This allows us to compare sessions according to  intensity.
After disaggregating data, we can more accurately measure the relationship between performance and practice. While performance generally improves with practice, we find that players tend to quit after an abnormally high score, 
suggesting significant rewards in casual games may instead encourage players to leave.
Interestingly, we find that players who are less likely to quit after a score drop tend to become more successful later.
Quitting is not as strongly correlated with skill, suggesting that it is perseverance to poor outcomes, i.e., grit, that contributes to player success.

To identify a plausible mechanism of game play, we train an $\epsilon$-machine, a type of a hidden Markov model, on the data, and find models that maximizes the accuracy of predicting players' performance. We find that players are most predictable when we model how their behavior is affected by changes in score from their previous game, instead of, for example, the change from their mean score. This model leads to insights not just in how players leave the game but the dynamics of performance as well.

The sheer size of behavioral data opens new avenues for the study of individual variability of human behavior. The empirical methods and models described in this paper can help game designers create more engaging games that keep people playing longer. Investigation of factors associated with disengagement and quitting could reduce churn rate, increase retention, and improve user experience in social media and mobile applications. More interestingly, the methods proposed here could pave the way to individual assessment and performance prediction from observed behavior.

\section{Background and Prior Work}
Axon is built to improve skill through engaged learning \cite{Ruben1999}, therefore it should come as no surprise that scores, a proxy for skill, increase with the number of games played, a proxy for practice~\cite{stafford2014tracing}, in agreement with studies of skill acquisition in laboratory and the field~\cite{ritter2001learning,newell1981mechanisms}.
In recent years, a more nuanced view of learning has emerged, one that emphasizes deliberate practice as a way to improve performance~\cite{Duckworth2011deliberate,ericsson2006influence}. Psychologists identified individual traits thought to be responsible for variations in the capacity for deliberate practice, such as \emph{grit}, which allows individuals to persevere in their efforts to achieve their long-term goals in the face of obstacles and challenges~\cite{duckworth2007grit}. While grit specifically refers to the ability to sustain efforts and passion for goals over extended periods of time, it is closely related to other psychological constructs such as persistence, resilience, conscientiousness, and self-control, which have been linked to achievement~\cite{Mischel89}. These traits may explain why some individuals have the self-discipline to continue practicing, even when faced with temporary setbacks, such as a drop in performance. While grit, as other psychological traits, is usually measured through surveys, identifying proxies of these traits, which can be computed from the observed individual behavior, has many practical benefits.

To better understand why users choose to persevere or quit, it is important to understand the psychology of motivation \cite{Locke2002,incentives}, especially the peak-end effect \cite{Fredrickson1993,Verhoef2004,Miran-Shatz2009,Gutwin2016peakend}, in which the individual's peak or last experience most affects their recall and motivation. Early work on goal-setting theory, e.g., suggests that moderate challenges encourage people to continue with a task, while extremely easy or difficult tasks reduce motivation. Other research, however, has found that the peak and end experiences change users' perception of the task \cite{Cockburn2015,Gutwin2016peakend,Fredrickson1993,Verhoef2004,Miran-Shatz2009}, which may then change their motivation to continue.
Other works have looked at how people play games, and use computers generally \cite{incentives,Gutwin2016peakend,Cockburn2015,Griffiths1993,Ryan2006}. For example, providing rewards increases the motivation for users to continue a task~\cite{incentives}, although the peak-end effect can sometimes affect user assessment of game difficulty and fun \cite{Gutwin2016peakend}, or otherwise affect user judgments \cite{Cockburn2015}.
Our paper hypothesizes that the peak-end effect may similarly help Axon game play.

Stafford \& Dewar~\cite{stafford2014tracing} empirically studied the impact of practice on performance using Axon game data. They examined the effects of \emph{practice amount} and \emph{practice spacing} on performance, and found that on average, game scores increased over consecutive plays, but there were significant differences between the higher scoring and lower scoring players.
The best performers had higher average scores than worse performers starting from the first games they played, and their score advantage grew with practice, in agreement with our paper's results.
Stafford \& Dewar also found that the longer the time period between the players' first and last games, the higher their scores are. Specifically, comparing players who played their first ten games within a 24-hour period with ``rested players'' who split their ten games over a longer period, they found that rested players had higher average scores than the former group.
They concluded that breaking up practice and resting, i.e., distributing practice~\cite{adams1987historical}, may benefit subsequent performance.
We show that after accounting for temporal structure of game play, this effect mostly disappears.

Our work differs significantly from, and expands on, work by Stafford \& Dewar. First, we split games into sessions of higher activity. As we show, this provides significant insight into the motivations users have to continue to play, such as stopping after a big score increase. Second, we find that scores increase significantly between sessions, but not between games, if we do not split them up into sessions, therefore the ``rested player'' result from Stafford \& Dewar may be due to peak score in the last game of a session and not rests between games. Furthermore, we find players who do particularly well in the game appear to exhibit grit, by refusing to quit when they perform poorly. 


Finally, we model users using a theoretically optimal and minimal HMM called an $\epsilon$-machine \cite{ShaliziCSMTheory}. It has been used recently to predict future user activity on social media \cite{Darmon2013,Darmon2015}, but is, to the best of our knowledge, a novel modeling framework in the field of computational psychology.

\section{Data and Methods}

\begin{figure}[t]
  \centering
\includegraphics[width=\columnwidth]{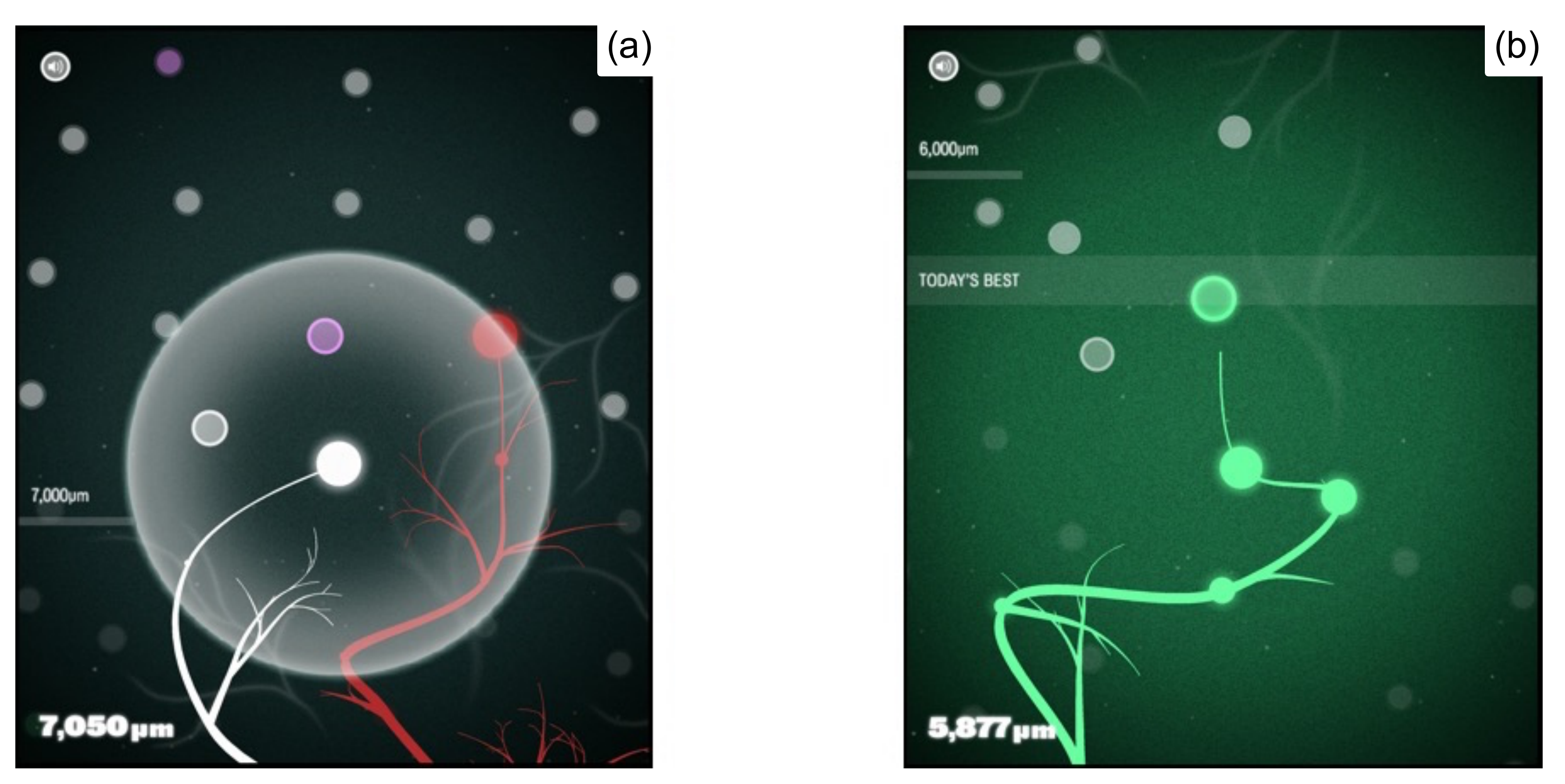}
  \caption{Screenshots of the Axon game showing (a) the player's neuron (white) and an enemy neuron (red), and (b) a power-up in effect.}\label{fig:axon-screenshots}
\end{figure}

The Axon game (http://axon.wellcomeapps.com) is a casual single player online game, where the player controls the growth of an axon 
(Figure~\ref{fig:axon-screenshots}). 
The game does not have levels of difficulty or time limits. Performance is characterized by a score that represents the length of the axon. Stochasticity is introduced in the game by ``power-ups'' (Figure~\ref{fig:axon-screenshots}b), 
which can significantly boost the score. 

\subsection{Data}
Stafford \& Dewar~\cite{stafford2014tracing} published the Axon game data at https://github.com/tomstafford/axongame. The data, collected between 14 March and 13 May, 2012, contains records of over 3M games played by more than 854K players. Each record contains the score and time of the  game (with hourly resolution), 
and a ``machine identifier'', an anonymized identifier derived from the web browser from where the game was accessed. Following Stafford \& Dewar, we assume that each machine identifier corresponds to a unique player. This need not be true, for instance, for shared computers or when a single player plays on multiple devices, but the size of the data is able to account for the noise produced by this phenomenon. The code used for our study is available at https://github.com/agarwalt/AxonGame.

The vast majority of people played only a few games: 92\% played fewer than eight games,
with 28K playing more than 12 games.
People who play few games may be systematically different from dedicated players who play many games; consequently, aggregating games across both groups can
lead to Simpson's paradox. To address this challenge, we segment each player's activity into sessions, where a session is a sequence of games without a long break (two hours or longer) between consecutive games. 


\subsection{Temporal Structure of Game Play}

\begin{figure}[t]
  \centering
\includegraphics[width=\columnwidth]{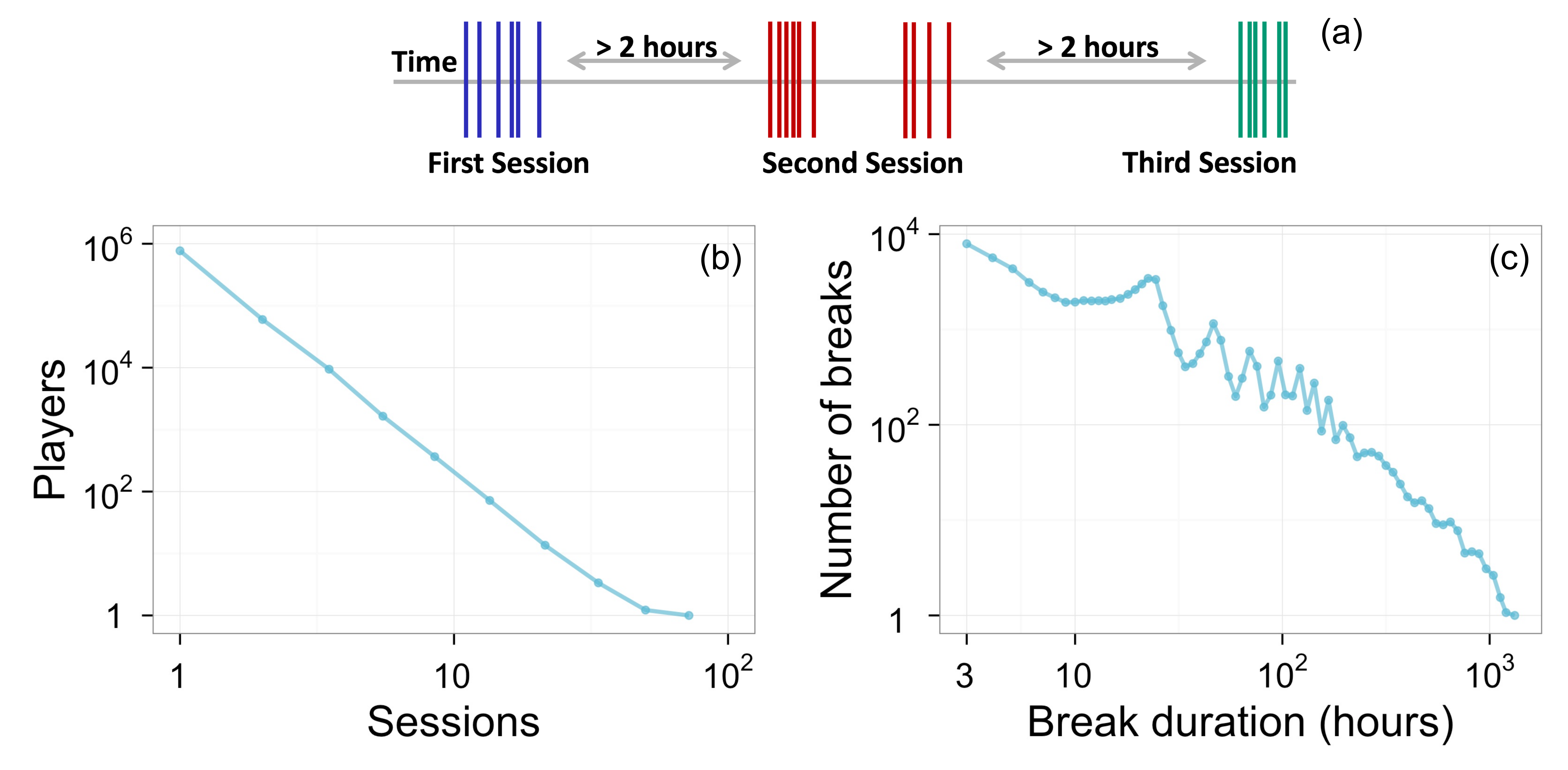}
  \caption{Sessions of game play. (a) Time series of games played by an individual, with colors denoting distinct sessions. A session can be arbitrarily long, and the time interval between consecutive games decides whether they belong to the same session. We use a threshold of two hours: any two consecutive games played more than two hours apart belong to different sessions.
  (b) Distribution of the number of sessions played, and (c) the time interval between consecutive sessions. Peaks are present at intervals of 24 hours.}\label{fig:players-vs-sessions-duration-between-sessions}
\end{figure}

People who play few games may be systematically different from dedicated players who play many games; consequently, aggregating games across both groups can
lead to Simpson's paradox. 
To address this challenge, we segment each player's activity into sessions, where a session is a sequence of games without a long break between consecutive games (Figure~\ref{fig:players-vs-sessions-duration-between-sessions}a). We use two hours as break threshold, but results do not change substantively when a different threshold, such as six hours, is used. Due to the long-tailed distribution of break time between consecutive games, changing the threshold affects only a small number of sessions. Segmenting player activity allows us to compare people who behave similarly, i.e., those who play similar number of games, rather than pool people with different behaviors together.

\begin{figure}[t]
  \centering
\includegraphics[width=\columnwidth]{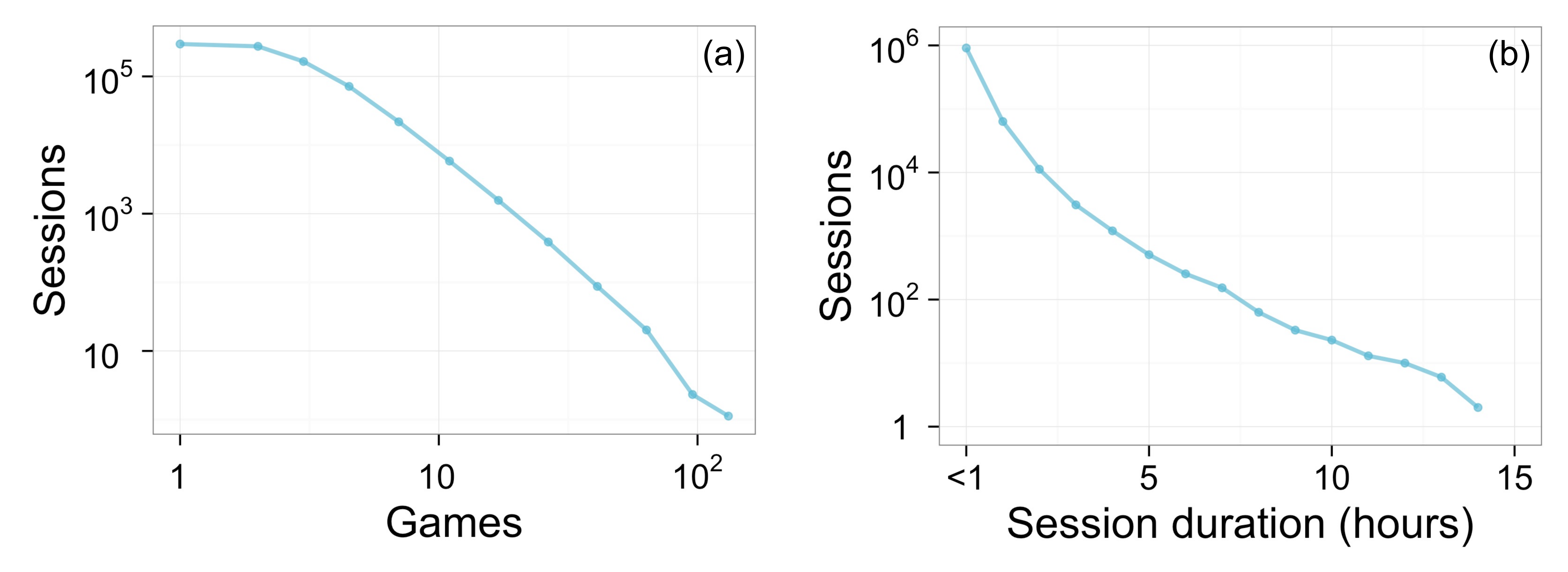}
  \caption{Distribution of (a) the number of games played per session and (b) the duration of sessions.}\label{fig:duration-of-sessions-games-per-session}
\end{figure}

Figure~\ref{fig:players-vs-sessions-duration-between-sessions}(b) shows the distribution of the number of sessions. Most players (about 90\%) have only one session, 
with the remaining 85K players who play more than one session. Daily and weekly peaks are present in the distribution of breaks between consecutive sessions (Figure~\ref{fig:players-vs-sessions-duration-between-sessions}c).
Of the 990K total sessions, more than 90\% last less than 1 hour (Figure~\ref{fig:duration-of-sessions-games-per-session}b), and 242K sessions played by 218K players have more than three games.
(Figure~\ref{fig:duration-of-sessions-games-per-session}a).

\subsection{Individual Variability}
Stafford \& Dewar found that the best and worst players, i.e., those who had achieved biggest and smallest personal high scores, respectively, differed in performance from the very first games they played. To partially control for this variability, we segment players by skill. We distinguish between two types of skill, which we refer to as \emph{talent} and \emph{success}. Talent is the initial skill of a player, which we operationalize as the median score of the first three games the individual ever played. Success is measured by the median of the three highest scores, or as many games as are available if less than three, after removing the first three, therefore we can only discuss the success of players who play four or more games. We follow tradition to use a player's best performance as a measure of success, rather than, for instance, an average value over time. Other measures of success, such as the median score over all triplets of consecutive games, are strongly correlated with our chosen criterion and do not change our conclusions. Further, using the median mitigates the effects of outliers.

\subsection{$\epsilon$-machine}
We model game play activity, 
including quitting the game and player's performance (scores),
using an $\epsilon$-machine, a type of Hidden Markov Model (HMM) with two useful properties. First, it is optimally predictive, meaning it produces the least uncertainly (entropy) about the future behavior of players~\cite{ShaliziCSMTheory}. Second, it is a minimally complex unifilar HMM, meaning that the model requires the fewest number of effective states, if transitions occur deterministically between states after each successive game (for a review of $\epsilon$-machines, see \cite{CrutchfieldReview}). In this paper, we fit an $\epsilon$-machine to our data using the Causal State Splitting Reconstruction (CSSR) algorithm~\cite{CSSR}. This algorithm groups past behaviors together into a single effective state if they make similar predictions. Importantly, this model is able to predict both score changes and the probability a user will quit the game, a property not possible with, for example, autoregressive models.

The $\epsilon$-machine, however, requires a few important assumptions to achieve these surprising feats. First, the model assumes a discrete time process that can be described with an alphabet. For continuous data, this means binning a real output into countable sets. 
Note that discretization results in information loss, and the machine will strongly depend on how the data is binned, 
but we discuss methods to address this issue in the next section. We define ``time" to be the game index, while the discrete alphabet is defined below.

Next, an $\epsilon$-machine assumes that the underlying sequence is stationary. This is clearly not true with respect to the absolute score (Figure~\ref{fig:score-vs-index-quartiles}), which tends to increase on average.
One way to approximate a stationary sequence, however, is to take the score difference between consecutive games.
We find that the difference is roughly independent of game index, as expected of a stationary distribution, until the final game before a user quits a session, when scores increase dramatically, while the quit rate itself is not strongly game index dependent.
We therefore create sequences of binned score differences; e.g., (0 to 7000 points better then the previous game), (over 7000 points better than the previous game), (quit);
which we model with an $\epsilon$-machine.


Finally, the $\epsilon$-machine assumes that we know the joint probabilities of the entire past and future of a sequence, but this is simply not practical.
Instead we allow for some error to enter the model, given a realistic amount of data.

\subsection{Evaluating Model Performance}

While binning data based on score differences leads to a roughly stationary sequence distribution, other potential sequences, such as binning by the difference between the current and mean scores, may be better modeled by the $\epsilon$-machine, and lead to more accurate predictions. Further, the sequences may be temporally correlated; past $L$ games may strongly affect future behavior, and we must pick a value of $L$ that is sufficiently long to produce accurate predictions, but not too long due to data limitations (and eventually computational cost).
In short, we must have a methodology for testing the goodness of the model.

The intuitive way to do this is to train the $\epsilon$-machine on a portion of data and test it on remaining portion, and determine whether we correctly discovered the next letter of our sequence using the model. By creating a cost function when our model creates an incorrect prediction, we can determine what values of $L$ or the alphabet improve predictions. Because our alphabet size $|\mathcal{A}|>2$, this becomes a multiclass classification problem. We can simplify this as a set of binary classification problems, however,
by predicting whether or not the next letter in the sequence is $X$, where $X$ is an arbitrary letter in our alphabet, where we train our model on $90\%$ of the data and test on the final $10\%$.

We apply a standard tool in binary classification problems:
the ROC curve, which tells us how often we correctly versus incorrectly predict a user quits, given a particular thresholding of the probabilities, e.g., $Pr(X)>p$, a user quits, otherwise they do not. We then calculate the area under the ROC curve (AUC),
which is equivalent to the Wilcoxon rank-sum test. Because of this equivalence, the limit variation in the AUC has a known form, however, we use bootstrapping of the testing data to non-parametrically find the errors of AUC values. Finally, we take the mean of AUC values across all $X$, weighted by the frequency of $X$ across all AUCs, in order to measure the model's overall efficacy in predicting players' scores or when they leave the game \cite{WeightedAUC}.

\section{Results}

\subsection{Success and Talent}

How much mastery of the game do players achieve? As defined earlier, a player's \emph{success} is the median score of his or her three best games after the first three games are removed (or as many as are available if less than three). A player's \emph{talent}, or initial skill in the game, is the median score of the first three games played. Overall, 
the correlation between success and talent for all players is $r=0.467$ ($p<10^{-6}$). 
Among players with the same skill, talent is only weakly correlated with success. For the initially least skilled players (bottom quartile by talent), the correlation between success and talent is $r=0.049$ ($p<10^{-6}$). The correlation for the second and third quartiles is $r=0.179$ and $r=0.137$, respectively ($p<10^{-6}$). The correlation is highest for players who are most skilled initially (top quartile by talent),  $r=0.299$ ($p<10^{-6}$). 

\subsection{Success and Practice} 

\begin{figure*}[t]
  \centering
\includegraphics[width=\textwidth]{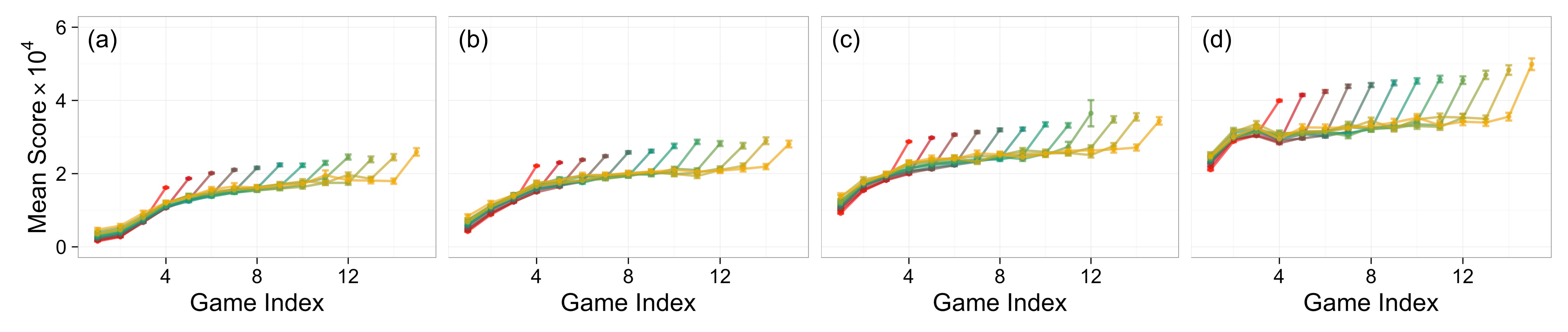}
  \caption{Score versus game play index for the (a) bottom, (b) second, (c) third, and (d) top quartiles by talent. Talent is measured by the median score of the first three games the player ever played. Lines represent sessions of different length, from 4 games to 15 games played in the session. Error bars represent standard error.}\label{fig:score-vs-index-quartiles}
\end{figure*}

How does practice---repeated game play---affect performance?
Figure~\ref{fig:score-vs-index-quartiles} shows the evolution of performance (average score) over the course of a session among players of similar skill (grouped by talent).
Lines represent sessions of different length, from 4 games to 15 games played in the session. 
There were 242K such sessions (out of the total 990K), representing 1.4M games, or approximately half of the total 3M games.

The figures reveal interesting trends. First, performance generally increases with the number of games played, reflecting the benefits of practice.
Second, eventual performance depends on skill: the most talented players (top quartile) have a better score, on average, on their very first game of a session than the least talented players (bottom quartile) have after practice. While the plot reflects performance averaged over all player sessions, these differences, also noted by Stafford \& Dewar, remain strong when only the player's first session is considered (data not shown). Finally, the very last game of a session has an abnormally high score, on average.
Aside from this last game, performance curves for sessions of different lengths within the same population overlap, suggesting that we properly captured the underlying behavior.


\begin{figure}[tb]
  \centering
\includegraphics[width=\columnwidth]{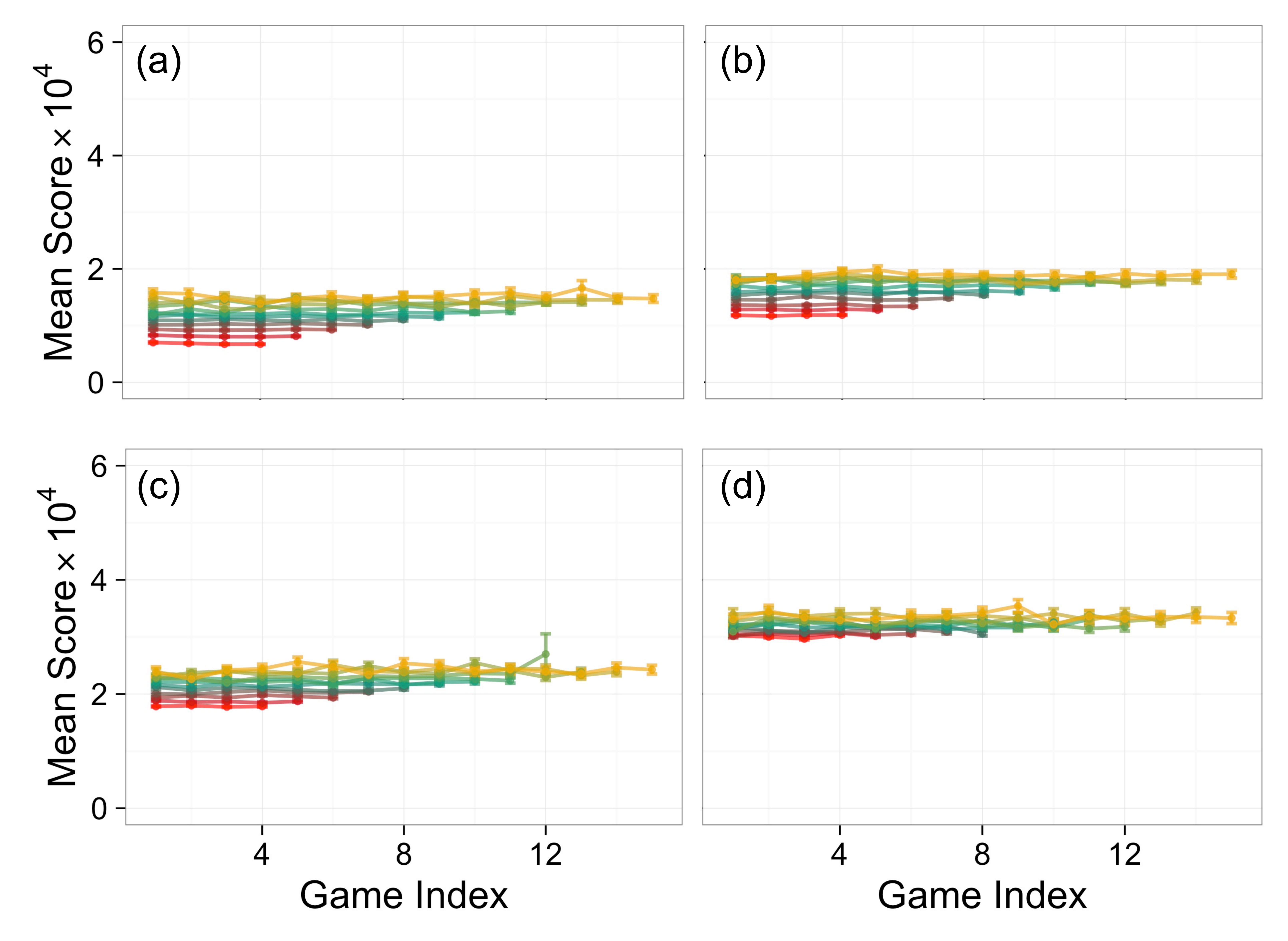}
  \caption{Score versus game play index for the (a) bottom, (b) second, (c) third, and (d) top quartiles by talent with games inside each session shuffled. Sessions (their length, positions, etc.) remain the same. Error bars show standard error. We notice that the distinctive upward score trend observed in the original data vanishes. If, on the other hand, \emph{sessions} are shuffled -- that is, the order of games is preserved but their timestamps are shuffled over all the games played by each player -- the trends are similar to those observed for the original data (not shown). This is probably because of coarse time stamps and the fact that most of the players (approximately 90\%) have only one 2-hour session.}\label{fig:score-vs-index-quartiles-shuffled-indices}
\end{figure}

To check robustness and rule out Simpson's paradox, we repeat the analysis on randomized data, where the indices of the games within each session are shuffled. Figure~\ref{fig:score-vs-index-quartiles-shuffled-indices} shows the resulting performance curves: performance no longer depends on the order the games within the session are played. 
This lends support for 
the claim that players' performance improves with practice. However, we cannot rule out alternate, if unlikely, explanations, such as the game designed to become progressively easier to play. 
Another notable observation about Fig.~\ref{fig:score-vs-index-quartiles-shuffled-indices} is that performance curves are stacked, with shorter sessions falling bellow longer sessions in terms of average scores. This suggests that players perform worse, on average, during shorter sessions than during longer ones.



The high score in the very last game in the session 
partly explains the performance boost that Stafford \& Dewar attributed to practice spacing. They found that players who split their first ten games over a time period longer than a day had higher scores on average than players who played their first ten games within the same 24-hour time period. We explain their observation differently.
Spacing the games over a time period longer than a day means that the player had to play the games over at least two sessions. Those who played their first ten games on the same day may have played multiple sessions, but are more likely to have played just one session.
Therefore, the higher average performance of the first group may be skewed by the high score of last game of the session compared to the second group. 

\begin{figure}[t]
  \centering
\includegraphics[width=\columnwidth]{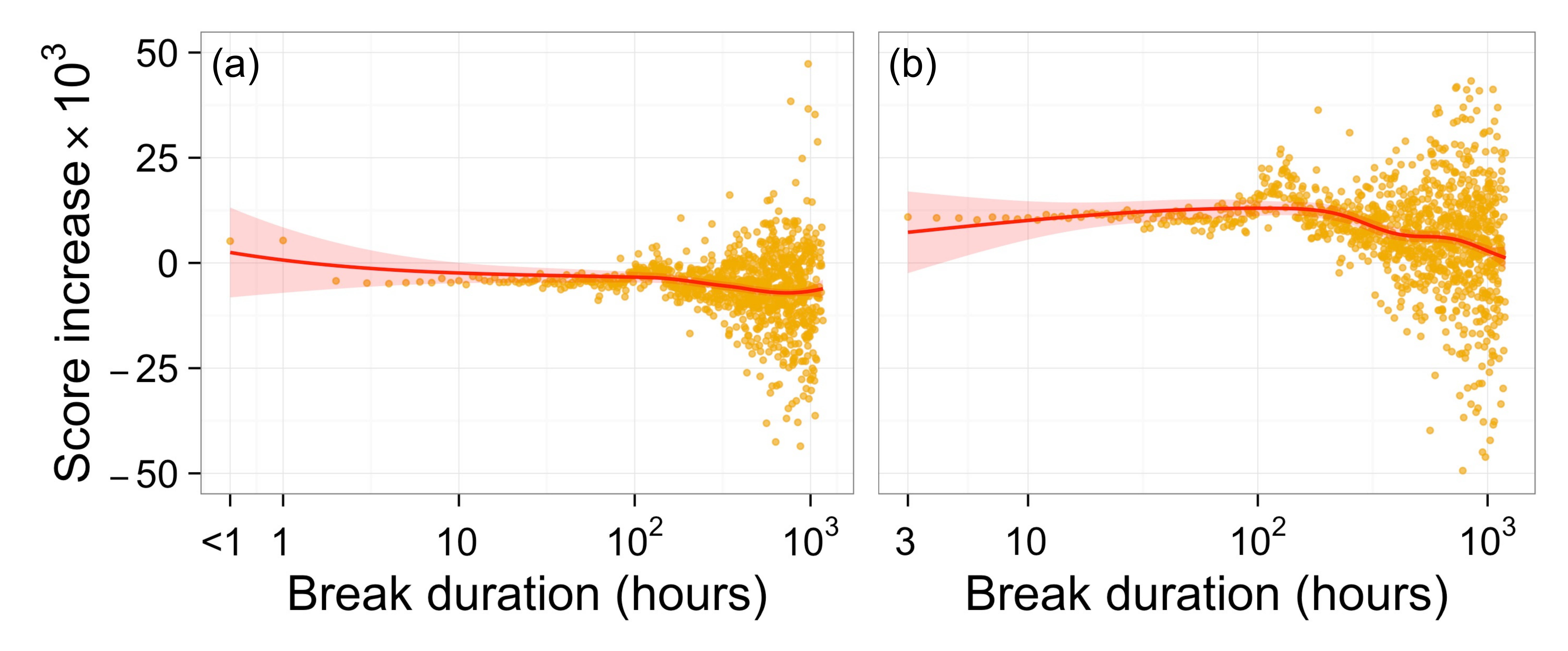}
  \caption{Average improvement in score versus length of the break between consecutive (a) games and (b) sessions. The shaded region represents 95\% confidence level in the non-parametric trend. A few outliers are omitted from the figure. Score improvement from one session to the next is defined as the difference between the median score of the first three games of the second session and median score of the last three games of the first session. The very last game of the first session is excluded from analysis. The main observations do not change if all games are included. }\label{fig:distributed-practice-games-and-sessions-temp}
\end{figure}

To explicitly measure the impact of practice spacing on performance, we plot the average change in performance as a function of the length of the break between two consecutive games and two consecutive sessions (Fig.~\ref{fig:distributed-practice-games-and-sessions-temp}). Change in performance between two games is simply the score difference between them. Change in performance between two sessions is the difference between the median score of the first three games of the next session and median score of the last three games of the previous session, where we exclude the last game of the session. 
Taking breaks between sessions does indeed lead to higher game scores (at the 95\% confidence level). 
However, the length of the break improves performance weakly for breaks less than a week; longer breaks result in smaller average improvement.

\subsection{Quitting}

Why does the last game of a session have a much higher score (on average)?
Do players simply choose to stop playing, thus ending the session, after receiving an abnormally high score?
To investigate this hypothesis, we empirically measure the probability to stop playing given the person played $n$ games.
We assume that this decision is based on a player's performance relative to his or her previous games. 
The relative performance can be measured as the difference from the mean or median score so far in the session, difference from the previous game's score, etc., but based on prediction accuracy, score difference from the previous game best models player behavior.

\begin{figure*}[tb]
  \centering
\includegraphics[width=\textwidth]{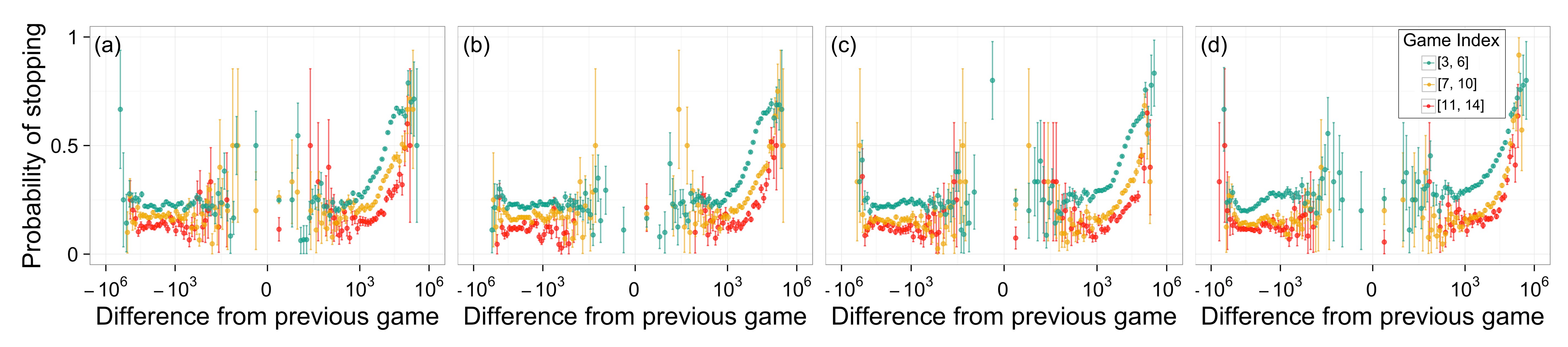}
  \caption{Probability to stop playing (thereby ending a session) versus score difference from the previous game for (a) the bottom, (b) second, (c) third, and (d) top quartiles of talent. Colors indicate over which indices the probability was calculated: $3-6$ (green), $7-10$ (yellow), or $11-14$ (red). Error bars are standard errors. We notice that the plots approximately overlap which suggests that the rate of quitting is nearly a stationary process, however at early indices, the quitting rate is higher than later ones.  
  }\label{fig:probability-stopping}
\end{figure*}

Figure~\ref{fig:probability-stopping} shows the quitting probability versus score difference from the previous game, $\Delta$, for different populations of players when split by talent. The quitting rate is simply the number of users who quit at score difference $\Delta$, divided by the number who ever reach $\Delta$ over a given range of game indices.
For 10K$< \Delta <$ 15K, players are more likely to stop playing
(Figure~\ref{fig:score-vs-index-quartiles}), even though large $\Delta$ does not correlate directly with any single game feature, such as power-ups. However, a concerted use of power-ups in succession can result in an increase of more than 10K points.  Surprisingly, for $\Delta < 0$, the quitting rate is not strongly dependent on $\Delta$ (Figure~\ref{fig:probability-stopping}). Should the designers of such games, then, avoid adding game elements which ``satisfy'' a player and potentially cause them to lose motivation to play? Answering this requires controlled experiments and is beyond the scope of this study.

\subsection{Success and Persistence} 

\begin{figure}[t]
  \centering
\includegraphics[width=0.7\columnwidth]{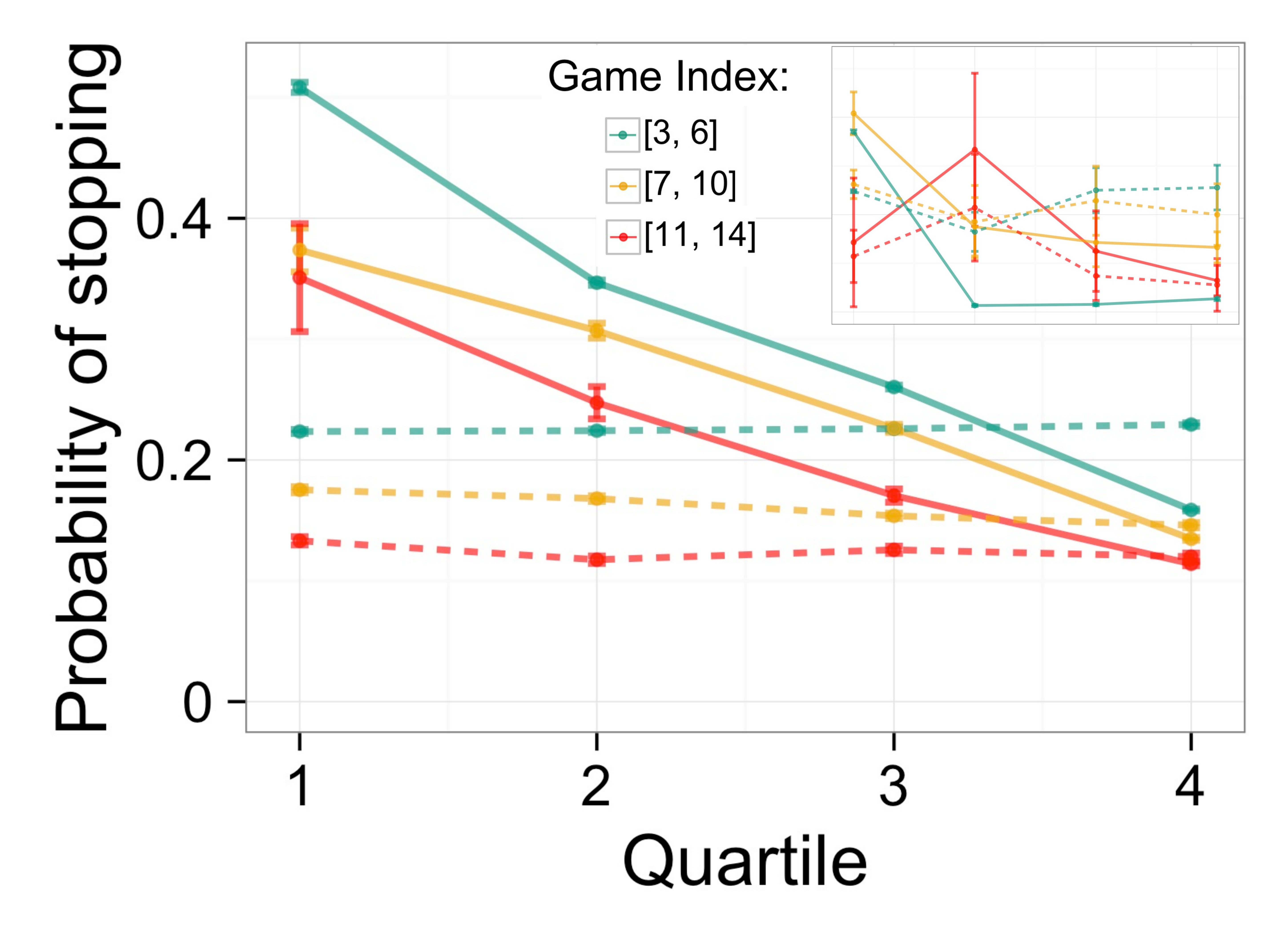}
  \caption{The probability to quit when scoring less than in the previous game. The lines represents quartiles of players split by success (solid line) and talent (dashed line). Error bars show standard error. The first quartile represents the least talented (successful) players and the 4th quartile the most talented (successful) players. The probability of ending a session -- the quitting probability -- is defined as the ratio of the number of games with scores less than the previous game's score which come at the end of session to the number of all games with scores less than the previous game's score.
  Inset: the probability to quit when scoring more than the previous game does not necessarily decrease for higher quartiles, which suggests that perseverance despite a low score best distinguishes successful players from unsuccessful players, rather than simply resistance towards ending a session.
  }\label{fig:quitting-vs-quartiles}
\end{figure}

Why do some people quit while others continue to play even when doing poorly (i.e., obtaining a worse score)? These \emph{persistent} players may possess a trait psychologists call \emph{grit}, which has been linked to high achievement and success~\cite{duckworth2007grit}. To investigate the impact of persistence on performance, we first need to quantify persistence, which we operationalize as the probability to stop playing after underperforming, i.e., obtaining a score less than the previous game's score. 
Figure~\ref{fig:quitting-vs-quartiles} shows the average persistence---or probability to quit playing after getting a worse score---for different quartiles of players as split by success or talent. Interestingly, we see a relationship between performance and persistence only in subpopulations of players segmented by success: the more successful players (those who achieve higher best scores) are less likely to stop playing after a setback, i.e., receiving a worse score.
In contrast, the relationship does not appear to be very strong when players are split by talent (their initial skill). Moreover, these trends do not hold in the probability to quit playing after receiving a better score in a game (inset in Figure~\ref{fig:quitting-vs-quartiles}).
This suggests that successful players are not simply ones who play longer; rather it is their ability to persevere despite lower scores that distinguishes them from less successful players.

Thus, consistent with psychology research on grit, persistence is associated with high performance and success, and not talent. Furthermore, successful players do not simply play longer; rather their ability to persevere despite lower scores distinguishes them from the less successful players.


\subsection{Modeling Performance}

%
%

Finally we ask: what mechanism best explains whether a player will quit or improve his or her score? To help answer this question, we create $\epsilon$-machines that model game play: how well users perform, and how likely they are to quit based on past performance. Our goal is to find the simplest, most parsimonious model with the highest predictive power. We choose a model with an alphabet of size four: one symbol denotes the state in which players quit the game (Q), while the remaining symbols denote performance that is ``poor", ``good", and ``very good".
We look at alternate ways to capture performance:
\begin{enumerate}
\item Score difference from the previous game,
\item Difference from a player's median score, and,
\item Difference from a player's mean score.
\end{enumerate}

In each case, binning the data into the states ``good" or ``very good" can be optimized by varying the bin size to  maximize model's performance. In addition, the $\epsilon$-machine can remember up to $L$ past games.
Due to the decay in the number of games any player plays, we vary $L$ only between one and three. To evaluate the model's performance, we create ROC curves from the predictions of each state, find the corresponding area under the curve (AUC), and then take the mean weighted by the frequency of each symbol across all AUCs, following work by Provost \& Domingos \cite{WeightedAUC}. To maximize the amount of training data the $\epsilon$-machine uses~\cite{CSSR}, we use $90\%$ of the data for training and reserve $10\%$ for testing.
We bootstrap the testing data to determine the confidence intervals of the AUC values.

\begin{figure}[t]
  \centering
\includegraphics[width=\columnwidth]{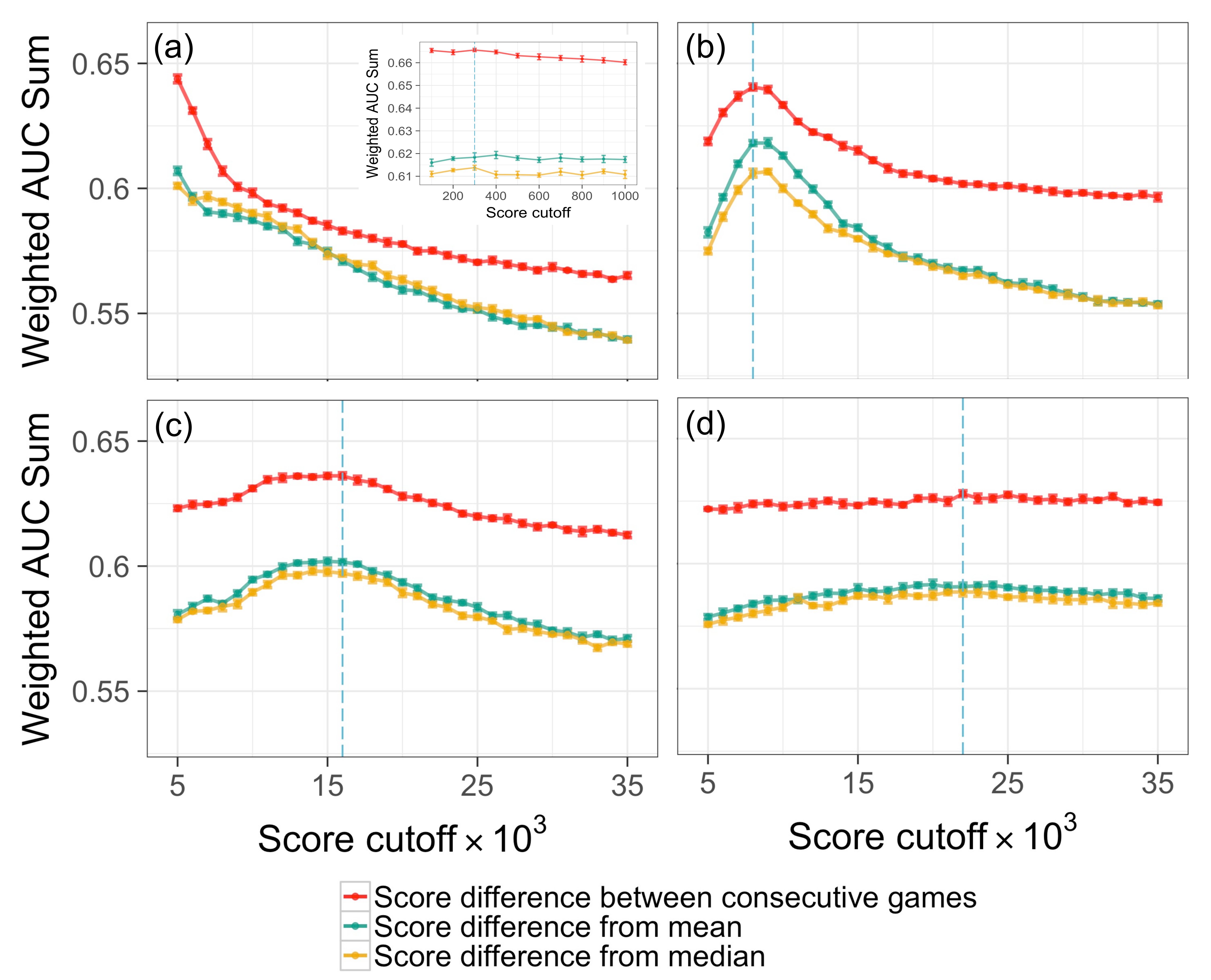}
  \caption{Weighted AUC values for different models and (a) bottom, (b) second, (c) third, and (d) top quartiles, split by talent. Alphabet sets are defined using score difference from mean or median score so far, or the score difference between consecutive games. The values on the x-axis are not the score differences themselves, but score difference cutoffs used to create the 4-letter alphabet. Error bars depict standard errors. The vertical dashed lines represent the approximate optimal score cutoff corresponding to the model on score difference between consecutive games.}\label{fig:auc-weighted-sum-all-quartiles}
\end{figure}

We trained the models separately on each quartile of players, split by talent. The results, shown in Figure~\ref{fig:auc-weighted-sum-all-quartiles}, 
suggest that score difference from a player's previous game ($\Delta$) leads to the best overall model, with an average AUC of roughly 0.64. Moreover, binning the data using thresholds $\Theta=300,~8K,~16K$, or $22K$ for each respective quartile maximizes prediction accuracy. Thus, a player can have ``poor" ($\Delta<0$), ``good" ($0\le\Delta<\Theta$), and ``very good" performance ($\Theta\le\Delta$).

\begin{figure}[t]
  \centering
\includegraphics[width=\columnwidth]{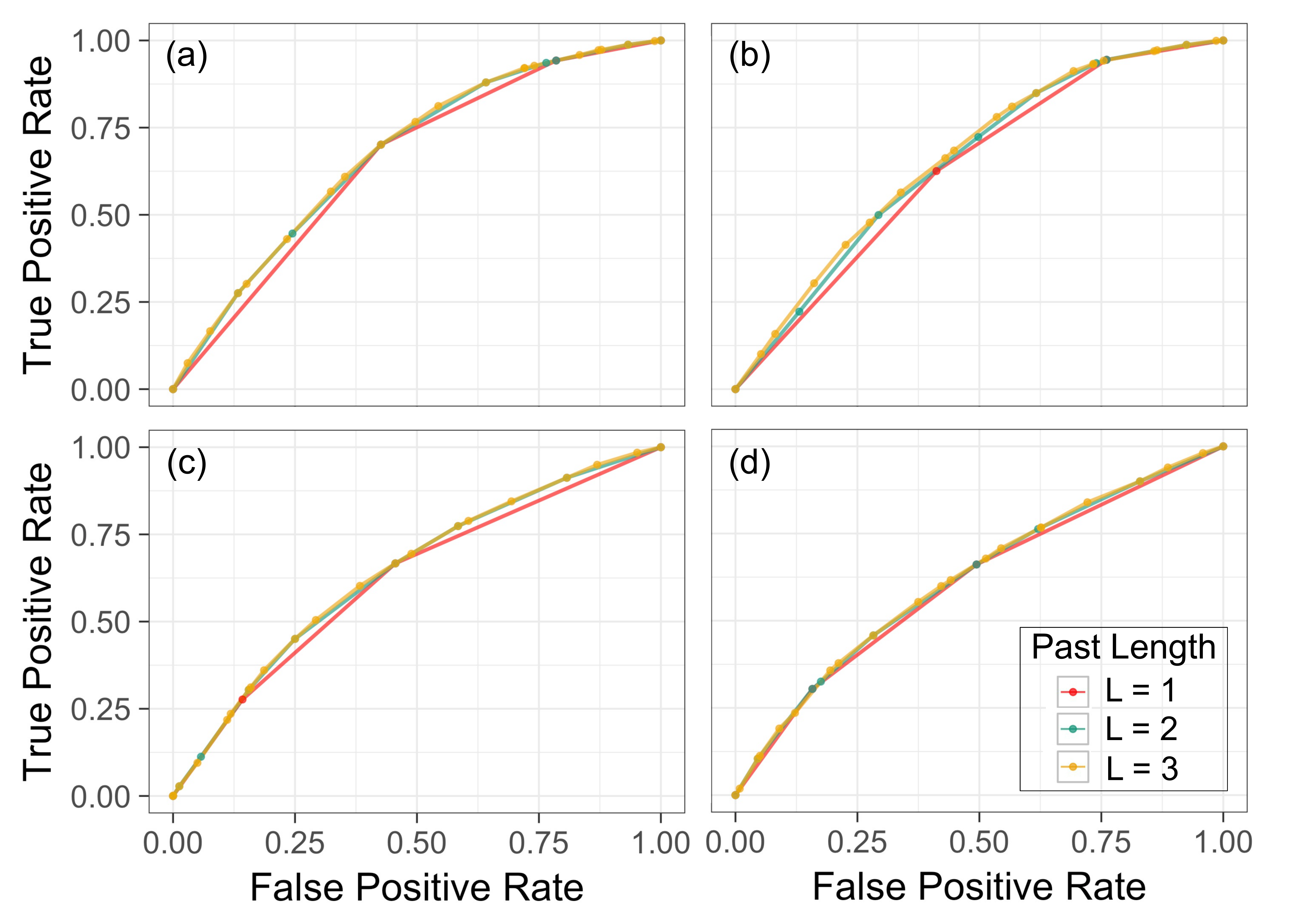}
   \caption{Effect of increasing past-length on ROC curves for predicting a player's behavior (shown are the ROCs for the fourth quartile, as a prototypical example). We focus on past-lengths of 1 (red), 2 (green), or 3 (yellow). We predict score changes $\Delta$ of (a) $ \Delta \le -1$, (b) $0 \le \Delta \le 21,999$, (c) $22K$ or more points, or (d) if they quit. 
}\label{fig:roc-increasing-history-lengths}
\end{figure}

We further test the importance of the past $L$ states, and find that a longer $L$ produces a higher AUC for sequences made using the score difference between consecutive games. To do so, we test and train on the 4th game index onward (255K out of 990K sessions) to make the ROC curves of different $L$s comparable However, when we study the ROC curve, we find that the points nearly coincide (see Fig.~\ref{fig:roc-increasing-history-lengths} for typical examples of ROC plots on the Axon game dataset), and much of the loss or gain in the AUC appears to be explained by lower $L$ models having fewer states and therefore fewer points on the curve. The near collapse of these curves may suggest that the most recent game has the strongest effect on user behavior. The number of states for longer $L$ models is also much larger (4 states for $L=1$ compared to 9 and 21 for $L=2$ and $L=3$, respectively, in Figure~\ref{fig:roc-increasing-history-lengths}). Therefore, we focus on the $L=1$ model.

\begin{figure}[t]
  \centering
    \includegraphics[width=0.7\columnwidth]{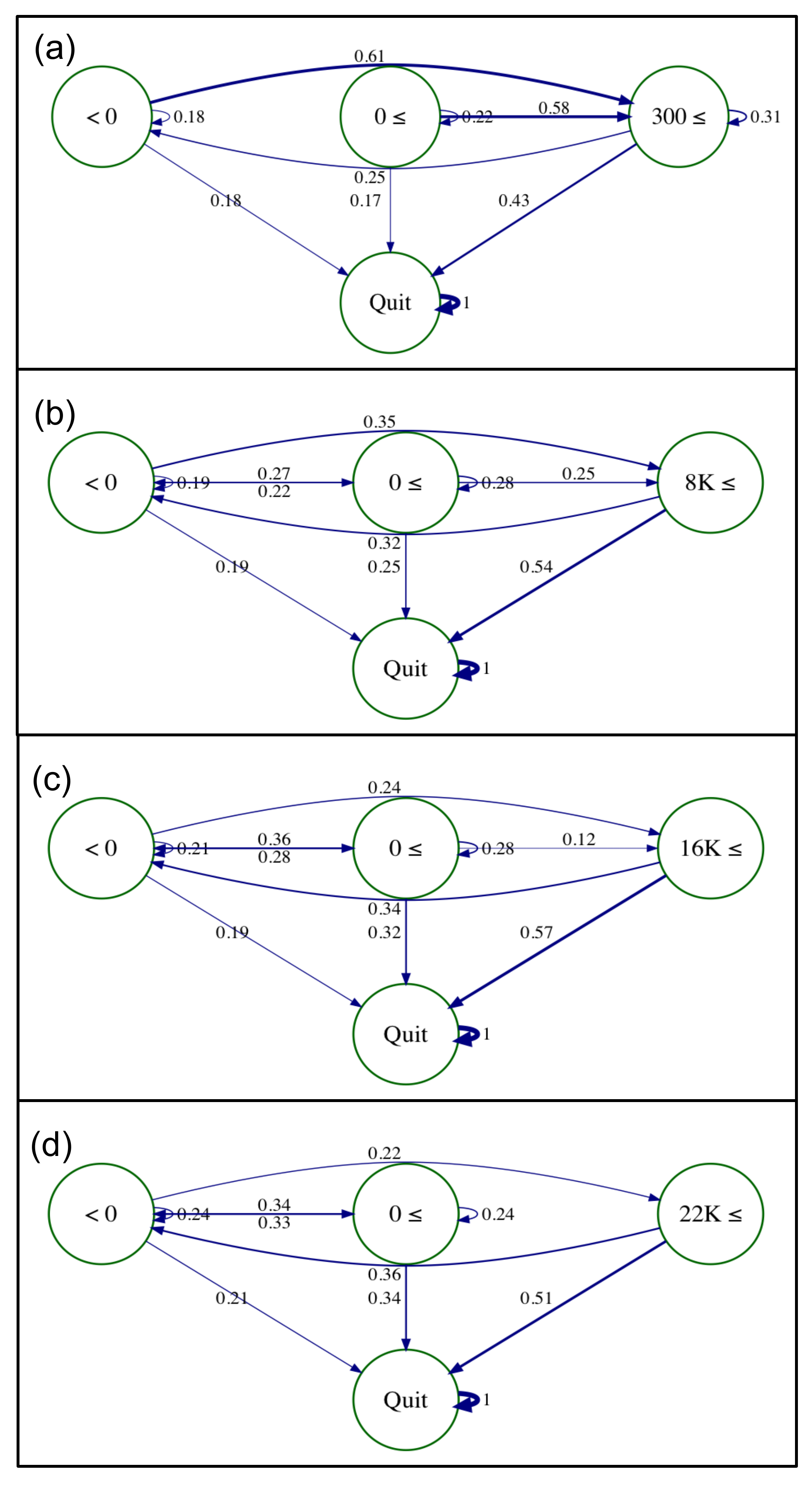}
  \caption{Diagrams of the optimal $\epsilon$-machines for players in the (a) first, (b) second, (c) third , and (d) fourth quartiles by talent. Thicker arrows represent higher transition probabilities, where a transition to a state occurs when the state's associated sequence appears in the data stream. Transitions with probability less than 0.1 are omitted for clarity.
  }\label{fig:state-diagram}
\end{figure}

The best overall models are shown in Figure~\ref{fig:state-diagram}, where the thickness of each line is proportional to the transition probability to a new state, such as ``poor" or ``good.'' Not only does the probability of quitting increase when the score increases over a few hundred to a few thousand points, in agreement with Figure~\ref{fig:probability-stopping}, but also that users transition in unexpected ways between states before they eventually quit. For example, players who perform poorly in a game (the change in score is negative) are very likely to perform well (reach the highest scoring state) in the next game. Similarly, there is an unexpected probability to transition from a ``very good" state in the last game to a ``poor" state in the next one, which suggests that players undergo periods of score volatility. Finally, the transition rates from negative to positive states are greater than the opposite transition rates in several quartiles, which suggests that players tend to improve over time.

\subsection{Confounding Factors}

Because this is an empirical study, there are a number of potential confounding factors, which may affect our conclusions. Previous work, e.g., has found that the high payout intervals (similar to frequent changes in scores), high self-reported skill, and even light or sound effects contribute to the motivation to play \cite{Griffiths1993}. Similarly, we do not measure Axon's perceived arousal, pleasure or dominance, which may explain why some users quickly quit playing and others do not regardless of their score \cite{Mehrabian1986}. Finally, we cannot determine to what degree exhaustion contributes to user behavior. For example, sessions of intense game play may exhaust gamers, and rest between sessions may not improve learning but instead allow users to improve mood, or replenish glucose levels, and therefore improve their score. Rest, independent of learning, is known to affect behavior in cognitively demanding tasks \cite{JudgeRuling}. Future work is necessary to correct for these effects.

\section{Conclusion}
We empirically investigated factors affecting performance in an online game using digital traces of activity of many players. The massive size of the data enabled us to investigate sources of individual variability in practice and performance.

Skilled (or talented) players, who score high already in their first games, are more successful overall. However, continued practice improves the scores of all players.
We identified a factor, related to grit, which captures the likelihood the player will keep practicing, i.e., playing the game, even when performing poorly. The more likely the player is to continue playing after 
a drop in performance, the more successful he or she eventually becomes. However, the ability to persevere and continue practicing is not related to player's initial skill.

We modeled this behavior using an $\epsilon$-machine and found 
that the model in which players based their decisions on how well they did compared to their previous game best predicted whether they will continue playing and their performance.
Surprisingly, when players did very well compared to their last game, they were highly likely to quit, 
but when they performed poorly, their quitting probability remained low. 

Our analysis relied on identifying and accounting for the sources of heterogeneity in game play data. Unless this is done, analysis can fall prey to Simpson's paradox, in which false trends can be observed when aggregating over heterogeneous populations.
Initial skill, or talent, is a major source of behavioral heterogeneity. Players who score well on their first games continue to improve and outperform the poorest players. A significant source of heterogeneity is the temporal structure of game play: players have periods, or sessions, of continuous activity with breaks in between. 
After accounting for sessions, a clearer picture of performance emerges.

While empirical analysis of behavioral data cannot replace controlled experiments, the sheer size of the data allows for the study of individual variability that is not possible with the smaller laboratory experiments. Such data can be used to explore alternate hypotheses about behavior, which can then be validated in the laboratory setting. Moreover, the types of quantitative methods explored in this paper could be used to predict performance and for psychological and cognitive assessment of individuals from their observed behavior. Future human-computer interfaces could continuously observe and predict our behavior, and adapt so as to optimize our performance.

\section{Acknowledgments}
This work was supported in part by NSF (\#SMA-1360058), ARO (\#W911NF-15-1-0142), and the USC Viterbi-India and ISI summer internship programs.

\small

\end{document}